\documentclass{emulateapj} 

\usepackage{amsmath}

\newcommand{\hateq}{=}
 
\begin{document} 
 
\title{The Fate of High-Velocity Clouds: Warm or Cold Cosmic Rain?} 
 
\author{Fabian Heitsch\altaffilmark{1}} 
\author{Mary E. Putman\altaffilmark{1,2}} 
\altaffiltext{1}{Dept. of Astronomy, University of Michigan, 500 Church St.,  
                 Ann Arbor, MI 48109-1042, U.S.A} 
\altaffiltext{2}{Dept. of Astronomy, Columbia University, 550 W 120th St.,  
                 New York, NY 10027, U.S.A} 
\lefthead{Heitsch \& Putman} 
\righthead{Fate of High-Velocity Clouds} 
 
\begin{abstract} 
We present two sets of grid-based hydrodynamical simulations of high-velocity clouds (HVCs) 
traveling through the diffuse, hot Galactic halo. These HI clouds have been suggested to 
provide fuel for ongoing star formation in the Galactic disk. 
The first set of models is best described as a wind-tunnel experiment in which the HVC
is exposed to a wind of constant density and velocity. In the second set of models we
follow the trajectory of the HVC on its way through an isothermal hydrostatic halo towards
the disk. Thus, we cover the two extremes of possible HVC trajectories.
The resulting cloud morphologies exhibit a pronounced head-tail structure, with a leading
dense cold core and a warm diffuse tail. Morphologies and velocity differences between
head and tail are
consistent with observations. For typical cloud velocities and halo densities, clouds
with H{\small{I}} masses $< 10^{4.5}$~M$_\odot$ will lose their H{\small{I}} content 
within $10$~kpc or less. Their remnants may contribute to a population of warm ionized
gas clouds in the hot coronal gas, and they may eventually be integrated in the warm
ionized Galactic disk. Some of the (still over-dense, but now slow) material
might recool, forming intermediate or low velocity clouds close to the Galactic disk.
Given our simulation parameters and the limitation set by numerical resolution, we
argue that the derived disruption distances are strong upper limits.
\end{abstract} 
 
%
%
\section{Motivation}\label{s:motivation} 
Galaxy disks require gas to fuel their ongoing star formation, yet
the fueling mechanism and the nature of the gas reservoir remain unclear.  
In galaxy formation models, the halo gas is initially heated as it  
follows the dark matter and subsequently cools and falls to the disk (e.g.,  
\citealp{1978MNRAS.183..341W}; \citealp{1991ApJ...379...52W}). 
Even at $z=0$, halo gas is proposed to gradually condense out  
into clouds, and a significant amount 
($\sim 4 \times 10^{10}$~M$_\odot$) of hot gas is thought to remain in an extended  
($\sim150$ kpc), hot, diffuse halo medium (\citealp{2004MNRAS.355..694M}; 
\citealp{2006ApJ...639..590F}; \citealp{2006MNRAS.370.1612K};  
\citealp{2006ApJ...644L...1S}). 
Other sources of potential fuel for galaxy disks include gas stripped  
from satellites, and possibly 
some mixed-in galactic fountain gas.   Understanding the  
properties of the diffuse 
hot halo and the clouds that fuel galaxy disks is a crucial step  
towards comprehending how galaxies form and evolve. 
 
   Observationally we know halo gas exists today around galaxies that is 
likely to fall towards the disk (v$_{gas} <$ v$_{esc}$) (e.g., 
\citealp{2001ApJ...562L..47F}; \citealp{2002AJ....123..873P};  
\citealp{2004ApJ...601L..39T}).  The neutral hydrogen halo clouds found  
around the Milky Way are called 
high-velocity clouds (HVCs) and are presumably a future source of star  
formation fuel for our Galaxy. 
HVCs are moving through a more diffuse halo medium 
as evident from their structure 
(e.g., \citealp{2000A&A...357..120B};   
\citealp{2003ApJ...597..948P}; \citealp{2007ApJ...656..907P}), 
and from detections of high-velocity O{\small{VI}} absorption  
\citep{2003ApJS..146..165S}. 
The clouds are disrupted by this interaction and form a ``head-tail''  
structure, i.e. the head of the cloud is compressed around a cold core, and a  
warmer, diffuse tail extends behind the cloud. The properties and  
lifetime of a given head-tail cloud depend on the density of the 
diffuse halo medium. 
Head-tail clouds therefore present the opportunity to study the  
lifetimes of halo clouds and 
the density of the diffuse Galactic halo, a medium that is 
difficult to detect directly.
 
In this paper we present three-dimensional grid simulations of HVCs
interacting with a diffuse halo medium and compare the results to 
neutral hydrogen observations. We discuss two sets of numerical
experiments, one best described as a wind-tunnel experiment in which
the model cloud is exposed to a wind of constant density and velocity, 
and a second set following the model cloud on its way through an
isothermal, hydrostatic halo towards the disk. All models include
equilibrium heating and cooling.
The characteristic timescales and pathlengths for disruption are assessed in the context 
of fueling the Galactic disk with the observed population of HI HVCs.
Our main results are summarized in Figure~6, predicting the distances HVCs of a given
mass are expected to survive their disruption at a given cloud velocity and background 
halo density.

%
%
\section{Technical Aspects}\label{s:technicals}

\subsection{The Numerical Scheme}\label{ss:proteus}
Calculations were performed with Proteus, 
an unsplit, second-order accurate gas-kinetic scheme 
\citep{1993JCoPh.109...53P,1999A&AS..139..199S,2001JCoPh.171..289X,2008ApJ...674..316H}
based on the Bhatnagar-Gross-Krook  
formalism \citep{1954PhRv...94..511B}. The scheme conserves mass, momentum and total 
energy to machine accuracy, and it allows the explicit control of viscosity and 
heat conduction, albeit at a Prandtl number of $1$ (see \citealp{2001JCoPh.171..289X}).
The dissipative constants are set such that they control the (very low) intrinsic
numerical dissipation (see \citealp{2002MNRAS.333..894S}, \S3 for a detailed description).

The thermal physics are implemented as a modification of the total energy via additional
heating and cooling terms. In the absence of heating and cooling, the gas is adiabatic with
an exponent of $\gamma=5/3$. The heating and cooling terms are accumulated in a 
tabulated ``cooling function'' (i.e. energy density change rate $\dot{e}$), depending 
on density $n$ and temperature $T$ as $\dot{e} = n\Gamma - n^2\Lambda(T)$. 
The energy change is done iteratively, i.e. if the local cooling timescale is 
smaller than the current Courant-Friedrich-Levy (CFL) timestep, we subcycle on the 
energy equation in that cell. This is generally the case in the cold dense gas.
Heating and cooling are operating throughout the whole simulation volume.
We derive the cooling function from the rates  
given by \citet{1995ApJ...443..152W} for $T<10^4$K, while for higher temperatures we use the rates of  
\citet{1993ApJS...88..253S}. We assume a metallicity of $10$\% solar  
\citep{1997ARA&A..35..217W,2001ApJS..136..463W}. 

The heating is due to a generic UV radiation field, which is set
to $G_0 = 1.1 X(z)$, where $X(z) = 1/[1+(z/8.53\mbox{kpc})^2]$ \citep{1995ApJ...453..673W}  
with $z=10$~kpc as a ``typical'' HVC distance. Although we do not know the actual distance to 
each of the head-tail clouds, we adopt $10$~kpc for most models based on the recent work constraining
the distances to many HVCs \citep{2006ApJ...638L..97T,2008ApJ...672..298W}. Many of the head-tail 
clouds are in the vicinity of larger complexes, suggesting that they may have a similar distance. 
This will be described further in the paper presenting the population of HIPASS head-tail clouds 
(Putman et al., in preparation). The UV-field is kept constant
during the evolution of the HVC. All heating and cooling processes are local, and we
do not address radiative transfer effects. Thus, the clouds cannot self-shield, and 
possible effects of increased UV radiation once the cloud approaches the Galactic plane
cannot be addressed. 

\subsection{Setup and Parameters}\label{ss:setup} 
We ran two sets of models. Series W are wind-tunnel experiments, while in series H, 
the cloud falls vertically towards the disk's midplane, traversing an isothermal halo medium in 
hydrostatic equilibrium. In both cases,
the clouds are initially at rest, located in the lower quarter of the simulation domain.

\begin{figure} 
  \begin{center} 
    \includegraphics[width=\columnwidth]{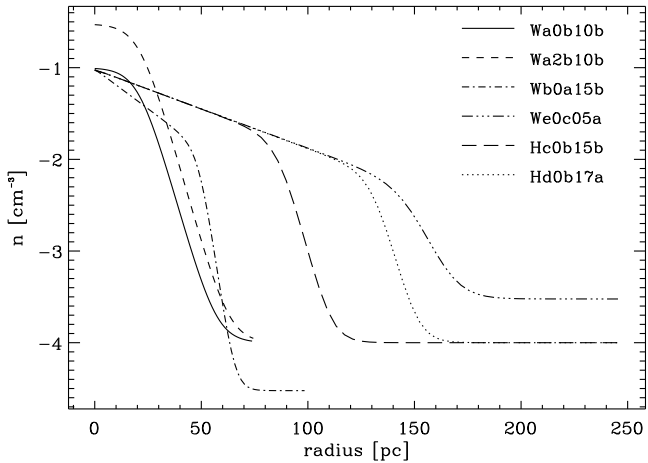} 
  \end{center} 
  \caption{\label{f:profiles}Logarithmic density profiles for models as indicated.
           The profiles are set by an exponential
           convolved with a $tanh$ shifted to the nominal cloud radius to make the cloud compact.}
\end{figure}

All model clouds have an exponential density profile (e.g. \citealp{2001A&A...369..616B}, see 
Figure~\ref{f:profiles}), convolved by a $tanh$ shifted to the nominal cloud radius to make 
the cloud compact. We also experimented with uniform density clouds and
found that they tend to disrupt even faster than the centrally peaked clouds, as expected.
Wind-tunnel experiments have simulation domain sizes varying between $150^2\times 300$ and
$500^2\times 1000$~pc$^3$, while the free-fall experiments use box sizes of $500^2\times 2000$~pc$^3$.

\subsubsection{Wind Tunnel Experiments}\label{ss:descW}
For series W, the background medium has a constant density $n_h$ and temperature $T_0$
such that the pressure is $100$~K~cm$^{-3}$ \citep{1995ApJ...453..673W} 
for models with $n_h \leq 10^{-4}$~cm$^{-3}$, and $300$~K~cm$^{-3}$ for models
with $n_h=3\times10^{-4}$~cm$^{-3}$ in order to avoid the peak of the cooling curve at 
$T\approx 3\times10^5$~K.
The cloud is initially in thermal pressure equilibrium with its surroundings.
To keep the cloud within the simulation domain as long as possible, we balance
a constant gravitational acceleration $\mathbf{g}\equiv g\hat{z}$ against 
a ``wind'' entering the bottom of the domain. The wind speed is ramped up linearly  
at a rate of $13$~km~s$^{-1}$~Myr$^{-1}$ until the desired ``cloud'' velocity $v_0$ is 
reached. The velocity $v_0$ sets the size of the gravitational acceleration $g$, via
\begin{equation}
M_cg = \pi R_c^2C_D\rho_h\,v_0^2/2,\label{e:terminal}
\end{equation}
with the cloud mass $M_c$, the drag coefficient 
$C_D=1.0$ and the halo density $\rho_h\equiv n_hm_H$.  
We note that while equation~(\ref{e:terminal})
implicitly assumes that the clouds reach a terminal velocity $v_0$, we use 
it only as a means to achieve an approximate balance
between gravity and ram pressure. It is not clear whether HVCs generally reach
their terminal velocities \citep{1997ApJ...481..764B,2007ApJ...656..907P}.
Moreover, equation~(\ref{e:terminal}) assumes that the clouds do not disintegrate,
leading to an over-estimate of the terminal velocity $v_0$. 
The lower $z$-boundary is defined by the ``wind'', while all other boundaries are 
left open.  

\subsubsection{Free-fall Experiments}\label{ss:descH}
For series H, we drop the HVC in a hydrostatic isothermal halo at $T_0=10^6$K. 
We use a gravitational acceleration of $g=10^{-8}$cm$^2$~s$^{-2}$, corresponding to the
fit between $1$ and $10$kpc provided by \citet{1997ApJ...481..764B}, and
we assume a halo density of $n_h=10^{-4}$~cm$^{-3}$ at $z=10$kpc. This results in a
density profile of 
\begin{equation}
  n_h(z) = n_0\exp(-g m_H/(k_BT_0))
\end{equation}
with a mid-plane density of $n_0=4\times10^{-3}$~cm$^{-3}$. The fact that this is lower than
the volume density of the ionized gas in the Galactic midplane by a factor of $5-10$
(see e.g. \citealp{2001RvMP...73.1031F}) is of no concern here, since our simple
model cannot describe the density structure for $z\lesssim 2$~kpc correctly anyway.
We will discuss in \S\ref{ss:interdist} how any systematic underestimate of the background halo
density affects the disruption of the clouds.
To keep the cloud within
the simulation domain, the grid is shifted by exactly one cell to lower $z$ 
once the center of the H{\small{I}} mass distribution 
has fallen at least by one cell. The new layer of cells at the bottom
of the domain is set according to the exponential density profile determined by
the above parameters. This prevents unnecessary and diffusive interpolation and it 
allows us to follow the cloud on its trip through the halo. 

Apart from this ``comoving'' grid, the major difference to series W is that the 
halo pressure is no longer constant. This leads to a successive compression of
the cloud and its fragments, until they drop below the resolution limit of the dynamical 
instabilities relevant for cloud disruption 
\citep{1997Icar..126..138R,2000Icar..146..387K}. This effect is amplified by the
increasingly higher cooling rates. In the extreme case, the remaining cloud
fragment can collapse to a thin ``needle'' in essential free-fall. 
To prevent such unphysical behavior, 
we regrid the simulation whenever the most massive fragment drops below the 
resolution limit. For such models, we simply reduce the domain size by one half, at the same
number of cells, thus doubling the spatial resolution. Typically, one regridding
step is necessary to follow the cloud evolution until disruption. It should be noted
that even with the regridding the smallest, cold dense fragments are not resolved with respect to the
hydrodynamical instabilities. Thus, their lifetimes tend to be over-estimated.

\subsubsection{Model Names}
The model parameters are summarized in Table~\ref{t:param}.  
The model names consist of seven characters, the first indicating whether the model
is a wind-tunnel experiment (W, \S\ref{ss:descW}) or a free-fall experiment (H, \S\ref{ss:descH}).
The second character denotes the cloud radius 
($a\hateq 21$, $b\hateq 50$, $c\hateq 89$, $d\hateq 133$, 
$e\hateq 150$~pc). The initial peak cloud density is given by the third character in $0.1$~cm$^{-3}$, 
while the fourth character indicates the halo density 
($a\hateq 3\times10^{-5}$, $b\hateq 1\times10^{-4}$, 
$c\hateq 3\times10^{-4}$~cm$^{-3}$). The fifth and sixth character stand for the nominal 
cloud velocity for wind-tunnel experiments, and for the peak cloud velocity for free-fall 
experiments, in $10$~km~s$^{-1}$. Finally, the seventh character stands for the resolution 
along the shorter axes ($a\hateq 128$, $b\hateq 256$~cells).

\begin{deluxetable*}{lcccccccccc} 
\tablewidth{0pt} 
\tablecaption{Model Parameters\label{t:param}} 
\tablehead{\colhead{Name}&
           \colhead{$N_x$}& 
           \colhead{$n_h$}& 
           \colhead{$n_c$}& 
           \colhead{$v_0$}& 
           \colhead{$M_0$}& 
           \colhead{$N_0$}& 
           \colhead{$R_0$} &
           \colhead{$Rn$} &
           \colhead{$\tau_0$}& 
           \colhead{$D_0$}\\
           \colhead{}& 
           \colhead{}& 
           \colhead{cm$^{-3}$}& 
           \colhead{cm$^{-3}$}& 
           \colhead{km~s$^{-1}$}& 
           \colhead{$10^3$M$_\odot$}& 
           \colhead{$10^{19}$cm$^{-2}$}&
           \colhead{pc}& 
           \colhead{cells} &
           \colhead{Myr}& 
           \colhead{kpc}}
\startdata 
Wa1b10b  & $256$  & $1\times10^{-4}$ & $0.1$ & $100$ & $1.4$ & $1.32$ & $21$  & $36$  & $35$ & $3.3$\\ 
Wa3b10b  & $256$  & $1\times10^{-4}$ & $0.3$ & $100$ & $4.5$ & $3.96$ & $21$  & $36$  & $65$ & $5.9$\\ 
Wb1a15a  & $128$  & $3\times10^{-5}$ & $0.1$ & $150$ & $5.9$ & $1.32$ & $50$  & $32$  & $50$ & $5.4$\\ 
Wb1a15b  & $256$  & $3\times10^{-5}$ & $0.1$ & $150$ & $3.0$ & $1.32$ & $50$  & $51$  & $59$ & $7.2$\\ 
We1c05a  & $128$  & $3\times10^{-4}$ & $0.1$ & $50$  & $44$  & $2.94$ & $150$ & $38$  & $150$& $5.3$\\ 
We1c05b  & $256$  & $3\times10^{-4}$ & $0.1$ & $50$  & $44$  & $2.94$ & $150$ & $77$  & $162$& $6.0$\\ 
We1c09a  & $128$  & $3\times10^{-4}$ & $0.1$ & $90$  & $44$  & $2.94$ & $150$ & $38$  & $81$ & $5.4$\\ 
We1b10a  & $128$  & $1\times10^{-4}$ & $0.1$ & $100$ & $44$  & $2.94$ & $150$ & $38$  & $64$ & $3.2$\\ 
Hc1b13a  & $128^a$& $1\times10^{-4}$ & $0.1$ & $130$ & $20$  & $2.55$ & $89$  & $46$  & $92$ & $9.0$\\ 
Hc1b15b  & $256$  & $1\times10^{-4}$ & $0.1$ & $150$ & $20$  & $2.55$ & $89$  & $92$  & $68$ & $9.2$ \\ 
Hc2b15a  & $128^a$& $1\times10^{-4}$ & $0.2$ & $150$ & $40$  & $5.10$ & $89$  & $46$  & $39$ & $6.7$ \\ 
Ha1b06a  & $128$  & $1\times10^{-4}$ & $0.2$ & $60$  & $7.3$ & $1.32$ & $21$  & $18$  & $36$ & $1.8$ \\ 
Hd1b17a  & $128^a$& $1\times10^{-4}$ & $0.1$ & $170$ & $37$  & $2.87$ & $133$ & $34$  & $103$& $11$ \\ 
Hd1b16b  & $256^a$& $1\times10^{-4}$ & $0.1$ & $160$ & $37$  & $2.87$ & $133$ & $68$  & $74$ & $11$ \\ 
Hd1b21a  & $128^a$& $1\times10^{-4}$ & $0.1$ & $210$ & $37$  & $2.87$ & $133$ & $34$  & $103$& $12$ 
\enddata 
\tablecomments{1st column: model name (see text),  
2nd: number of grid cells along shorter domain axes ($a\hateq 128$, $b\hateq 256$~cells), 
3rd: background halo density ($a\hateq 3\times10^{-5}$, $b\hateq 1\times10^{-4}$, 
$c\hateq 3\times10^{-4}$~cm$^{-3}$),   
4th: cloud peak density,
5th: nominal cloud velocity,  
6th: initial cloud mass, 
7th: initial peak column density, 
8th: initial cloud radius ($a\hateq 21$, $b\hateq 50$, $c\hateq 89$, $d\hateq 133$, 
$e\hateq 150$~pc), 
9th: resolution elements per cloud radius,
10th: characteristic life time (eq.~[\ref{e:rate_tau}]),
11th: characteristic distance traveled (eq.~[\ref{e:rate_dist}]).\\
$^a\,$ These models have been regridded (see \S\ref{ss:descW}).
}
\end{deluxetable*}

%
%
\section{Results} 
We start our discussion of the results in terms of morphology (\S\ref{ss:morph}), including 
a comparison to observed HVCs. A detailed view of three representative models is given in
\S\ref{ss:details}. We also summarize the cloud life times and their travel distances 
(\S\ref{ss:disrupt}), discuss resolution effects (\S\ref{ss:resoissue}), and differences to previous
work (\S\ref{ss:previous}, see also \S\ref{sss:increase}).

\subsection{Morphologies}\label{ss:morph}
\begin{figure*} 
  \begin{center} 
  \includegraphics[width=\textwidth]{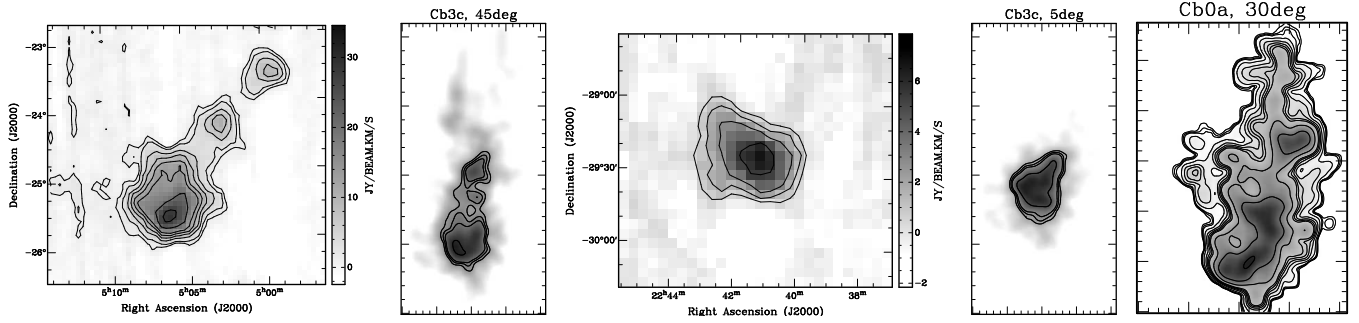} 
  \end{center} 
  \caption{\label{f:acviews}Column density maps of observed and modeled HVCs. Observed clouds 
           were taken from the HIPASS catalogue by \citet{2002AJ....123..873P}. From left to right: 
           HIPASS cloud \#648, model cloud Wb1a15b at $45$~Myr, rotated by $45^\circ$, HIPASS cloud \#200,  
           model cloud Wb1a15b at $45$~Myr, rotated by $5^\circ$, and model cloud We1c05b at $110$~Myr, 
           rotated by $30^\circ$. 
           Contours indicate column densities levels identical for all clouds, at  
           $(1.2,1.6,2,4,8,12,16,20)\times 10^{18}$~cm$^{-2}$. Because of the higher peak 
           densities, model We1c05b has 
           additional contour levels at $(40,80,160)\times 10^{18}$~cm$^{-2}$.
           The models have a nominal resolution of $\approx 2$~pc, corresponding to  
           $0.7'$ at $10$~kpc. The maps have been degraded by a Gaussian 
           corresponding to an angular resolution  of $3.0'$ for a more realistic comparison to observations.} 
\end{figure*}

Figure~\ref{f:acviews} shows column density maps of HI (i.e. gas at $T<10^4$~K for the model clouds) of 
observed and modeled HVCs with the same column density contours for each. 
Observed clouds were taken from the HI Parkes All-Sky Survey (HIPASS) cloud catalogue  
\citep{2002AJ....123..873P}.   Clouds with a head-tail structure were found for $\sim40$\% of the small
HVCs (CHVCs and :HVCs) in this catalogue (Putman et al., in prep.), and also along the Magellanic
stream \citep{2003ApJ...586..170P}.  \citet{2000A&A...357..120B} 
put together a flux-limited sample of head-tail clouds
with the Leiden-Dwingeloo Survey (35\arcmin~resolution vs. 15.5\arcmin~for HIPASS)
and found 20\% of their entire sample show a velocity and column density gradient consistent with
being head-tail clouds.
The modeled clouds in Figure~\ref{f:acviews} have been degraded to a ``beam width'' of
$3.0'$, and they have been rotated by the amount indicated on top of each 
model map. For small rotation angles, the cloud is traveling nearly head-on towards the observer.
Model Wb1a15b at $45^{\circ}$ develops a pronounced head-tail structure 
reminiscent of the structures in the observed cloud, including several fragments in 
the tail. Shear flow instabilities act along the cloud and  
tail flanks, and the tail itself shows kinks as a result of a turbulent wake.  
Cooling leads to fragmentation, resulting in ``multiple cores'' along the tail. 
In the head-on view (cloud Wb1a15b, $5^\circ$), the tail is nearly completely 
suppressed, but might be discernible in a velocity shift between the cold and warm gas component.  
We will explore the kinematic signatures of the clouds and their disruption in a subsequent paper. 
The panel on the right (model We1c05b, $30^\circ$) shows a more massive cloud at a late evolutionary 
stage. The cloud has disintegrated into multiple cores. The minimum 
temperatures in the cores reach $\approx 100$~K.

\subsection{Cloud Evolution and Disruption Mechanisms}\label{ss:details}
\begin{figure*} 
  \begin{center} 
  \includegraphics[width=\textwidth]{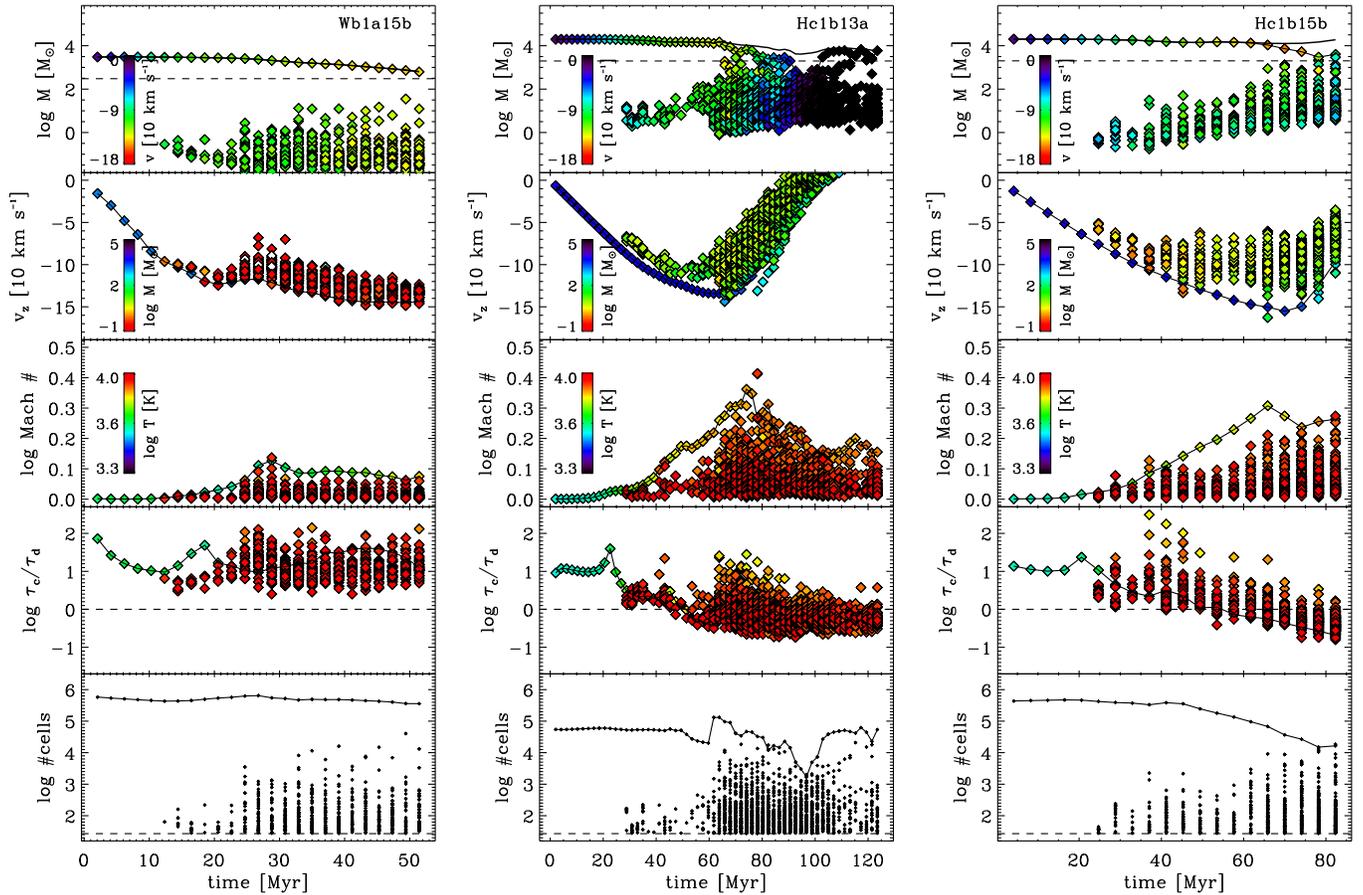} 
  \end{center} 
  \caption{\label{f:timesum}Time histories for models Wb1a15b, Hc1b13a and (at twice the resolution) Hc1b15b. 
           From top to bottom:
           logarithm of H{\small{I}} mass of all cloud fragments, their velocity, the internal Mach number of each 
           cloud fragment, the ratio of cooling time scale over dynamical timescale to estimate the 
           likelihood for disruption, and the number of cells in each fragment. Color codes as indicated
           in the panels. {\em Note that the velocities are given in $10$~km~s$^{-1}$.}
           The ratio $\tau_c/\tau_d$ is color-coded with the temperature as in the Mach number
           plot. Solid lines denote the evolution of the most massive fragment (i.e. the main body of the 
           cloud). For the mass-history, a thick solid line shows the total H{\small{I}} mass. The 
           dashed line at the bottom of the cell number plot indicates the threshold of $64$ particles 
           required to be accepted as a ``cloud''. Note that this is far below the resolution limit 
           for hydrodynamical instabilities (see discussion in \S\ref{ss:resoissue}).}
\end{figure*}

We will discuss the cloud evolution in terms of three models, namely Wb1a15b, Hc1b13a and (at twice
the resolution, but otherwise identical parameters) Hc1b15b,
representing windtunnel and free-fall models, and demonstrating resolution effects as well as long-term
evolution. Figure~\ref{f:timesum} summarizes the time history of the three models, from top to
bottom the mass, velocity, Mach number, ratio of thermal over dynamical time scales
and the number of grid cells against time. We will discuss each of them in turn.
To measure the time histories of the clouds, we ran a core finder identifying
coherent structures in three dimensions  with temperatures $T<10^4$~K as a proxy for neutral 
hydrogen. Structures with fewer than $64$ cells are not counted. In the following, we will use 
"cloud mass" synonymously with "cloud mass in H{\small{I}}".

\subsubsection{Mass evolution}
We begin our discussion with the mass histories (top panels of Fig.~\ref{f:timesum}),
which indicate the survival chances of the model cloud. The simulations start out with one 
cold cloud, which subsequently fragments due to a combination of dynamical and thermal
instabilities. The thick solid line
denotes the total mass in H{\small{I}}, while the thin solid line follows the mass evolution of 
the most massive fragment (usually the main cloud except for very late stages of nearly
complete fragmentation). The dashed line indicates $10$\% of the initial cloud mass. 
The center-of-mass velocity of each fragment is shown in color. 

With increasing time, more and more material is ablated from the cloud, resulting in a 
growing number of cloud fragments.
The increasing lower mass envelope of the fragments in model Hc1b15b (to some extent
visible in Hc1b13a as well) results from the recooling of the fragments.
With decreasing $z$, the cloud fragments are being compressed by the increasing background pressure,
thus leading to higher cooling rates and ``re-formation'' of cold H{\small{I}}-cloudlets. 
The mass fraction in the small, ``re-formed'' cloudlets amounts to $\approx 10$\% of
the H{\small{I}}-mass of the original cloud.
This effect is absent in the wind-tunnel experiments, where the cloud is being shredded into 
successively smaller fragments. The evolution of "real" HVCs we expect to lie in between 
these two extremes of exponential density increase and constant density of the
background medium, since
the cloud trajectories will
generally not be restricted to motions along $z$, specifically if the clouds stem from material
stripped off satellites. 

\subsubsection{Peak Velocities}
The second row of panels in Figure~\ref{f:timesum} complements the mass histories discussed
above. Although this information may seem redundant, it is easier
to glean the peak velocities reached, and the velocity distribution of the fragments.
As in the mass histories, the thin solid line traces the evolution of the most massive fragment.

The velocities of the cloud fragments are generally smaller than that of the main body, indicating
a growing ``tail'' of the cloud, due to dynamical drag. The lag is consistent with the 
velocity gradients in observed HVCs (\citealp{2000A&A...357..120B,2001A&A...370L..26B}; 
see also \citealp{2005A&A...432..937W}).

While for the wind-tunnel experiment, the velocity is prescribed, the velocities reached by the 
H-series range well below ballistic velocities (see e.g. \citealp{1997ApJ...481..764B}), and rarely exceed 
$\approx 150$~km~s$^{-1}$. Several effects play a role here. First, because the clouds are deformed
due to hydrodynamical interaction with the background gas, the drag forces might be higher than
expected. Second, the clouds lose mass and eventually disintegrate, and third, the increasing
background density leads to stronger drag. 
In fact, for model Hc1b13a, the combination of mass loss and background density increase brings the 
cloud fragments to a halt at around $100$~Myr. Together with the recooling of cloudlets (see previous section),
this offers an interesting possibility of reforming H{\small{I}}-clouds close to the disk (\S\ref{ss:feeding}).

Observationally, relative to the Galactic standard of rest (GSR), the maximum HVC velocities are 
$\pm 250$~km~s$^{-1}$, with the more typical GSR velocities $<100$~km~s$^{-1}$ in magnitude 
and a median value of $-30$~km~s$^{-1}$ (e.g. \citealp{1991A&A...250..509W}; \citealp{2002AJ....123..873P}). 
Though we do not know the 
true velocities of the clouds, this represents our only observational constraint. Many people
have begun to use the deviation velocity, or how much the HVC motions differ from a simple model
of Galactic rotation, when discussing HVCs (e.g. \citealp{2004ASSL..312...25W}). The majority
of the deviation velocities are $<200$~km~s$^{-1}$. The peak values of $\pm 300$~km~s$^{-1}$
belong to HVCs associated with the Magellanic System.

\subsubsection{Internal Mach Numbers}
The third row of Figure~\ref{f:timesum} shows the logarithm of the internal Mach number for all 
identified fragments.
We define the Mach number as the density-weighted rms velocity within a fragment over its
average sound speed. As an estimate of the sound speed, we show the volume-averaged temperature
of each fragment in color. Note that the lowest temperatures can reach $100$K in our models,
in agreement with observed values \citep{2004ASSL..312..145S}. 

The Mach numbers reach values up to $2$, although it should be noted that these are averaged
over the cloudlets -- in the coldest regions, the Mach numbers can exceed values of $5$. 
Thus, the clouds have strong internal dynamics, indicating dynamical instabilities at work.
Yet the Mach numbers and disruption timescales do not correlate -- for
that our measures at this point are too crude, and we defer a more detailed discussion to a future
paper. 

\subsubsection{Thermal over Dynamical Timescales}
The ratio of the cooling time over the dynamical timescale $\tau_c/\tau_d$ can be used as a measure
for the importance of thermal over dynamical instabilities. Cooling can stabilize the cloud against 
disruption by damping the propagation of waves into the cloud (e.g. \citealp{1997ApJ...483..262V}).
The cooling timescale is given by $\tau_c\equiv k_BT/(n\Lambda)$,
and as the dynamical timescale we take the sound crossing time $D/c_s$, where the diameter $D$ is estimated 
by taking the geometric mean of a cloud fragment's extent along the three grid axes, i.e. $D\equiv (D_xD_yD_z)^{1/3}$. 
For $\tau_c/\tau_d>1$, the cooling time is longer than the sound crossing time, and perturbations will
only be slightly damped when traveling into the cloud. For $\tau_c/\tau_d<1$, thermal effects dominate, 
leading to a stabilization of the cloud. If the densities and temperatures of the cloud fragments
were constant with time, 
dynamical effects should eventually win over thermal ones with progressive fragmentation
of the cloud. 

Cloud Wb1a15b is completely dominated by dynamics ($\tau_c/\tau_d > 1$ for all $t$). This is consistent 
with the gradual disruption of the cloud mirrored by the continuous mass loss (top panel). 
As the number of cells in the most massive fragment (bottom panel) stays nearly constant, 
the mean cloud density decreases, thus further lowering the importance of thermal effects.

Clouds Hc1b13a and Hc1b15b are initially dominated by dynamics, however, the increasing 
background pressure compresses the cloud and thus shortens the cooling time, rendering some of 
the fragments and eventually the main body of the cloud thermally dominated for $t\gtrsim 45$~Myr. 
Yet the cloud still disrupts because the dense gas enters the thermally unstable regime of 
the cooling curve. Fragmentation is enhanced since the condensation mode of the thermal instability 
has an outer length scale set by the product of the sound crossing time and the cooling time
\citep{2000ApJ...537..270B,2008ApJ...689..290H}.

\subsubsection{Cloud Resolution}
The bottom panels of Figure~\ref{f:timesum} show the number of cells in each cloud (fragment). The dashed
line denotes a limit of $64$ cells below which a fragment is not counted as a cloudlet any more. Clearly,
this number is well below the hydrodynamical resolution limit (see discussion in \S\ref{ss:resoissue}).
For the vast majority of the time, the main body of the clouds stays above $10^4$ cells. The jump at $60$~Myr in model
Hc1b13a is due to the regridding of the model (see \S\ref{ss:descW}).

\subsubsection{Cloud Travel Distances}\label{ss:traveldistance}
\begin{figure*} 
  \begin{center} 
  \includegraphics[width=\textwidth]{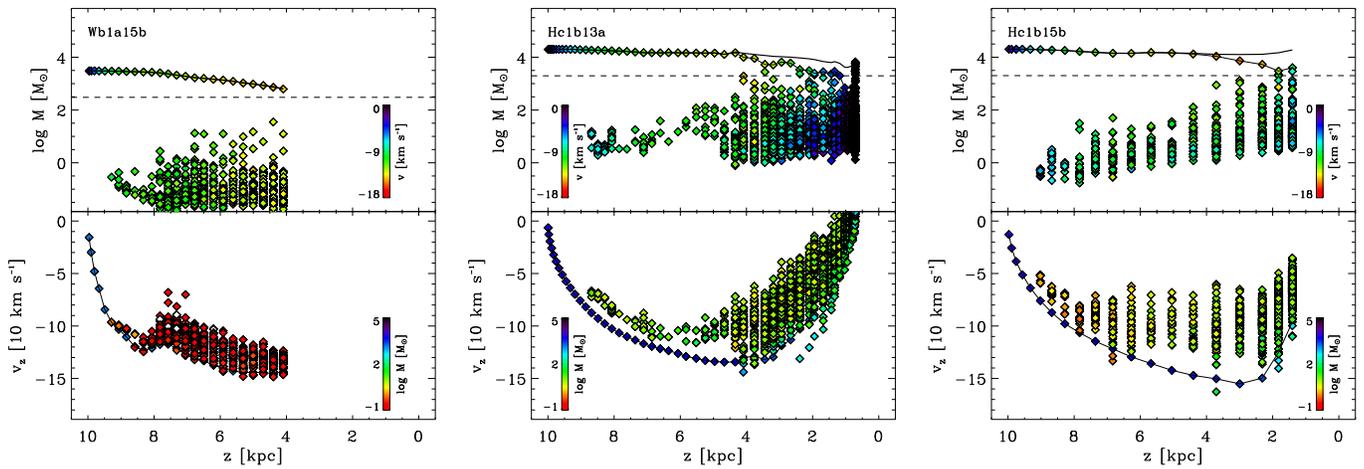} 
  \end{center} 
  \caption{\label{f:distsum}Mass (top) and velocity (bottom) of cloud fragments against distance from Galactic plane for
           models Wb1a15b, Hc1b13a and Hc1b15b. {\em Note that the velocities are given in $10$~km~s$^{-1}$.}
           Solid lines denote the evolution of the most massive fragment (i.e. the main body of the 
           cloud). For the mass-history, a thick solid line shows the total H{\small{I}}-mass. 
           Because of the small mass contained in the fragments of model Wb1a15b, the lines are
           indistinguishable for that model.}
\end{figure*}

Figure~\ref{f:distsum} shows the mass and velocity history in terms of the cloud's position $z$ above the 
Galactic plane, for models Wb1a15b, Hc1b13a and Hc1b15b. Strictly, for model Wb1a15b, the initial position
is arbitrary. Also note that the cloud velocities for the W-series are not self-consistent, but are imposed
by the wind entering the simulation domain (see \S\ref{ss:descW}). For the H-series, the velocity profiles of the 
most massive fragment (i.e. the main body of the cloud) are initially similar to those derived by 
\citet{1997ApJ...481..764B} (their Fig.~3) for clouds with a column density of $10^{19}$~cm$^{-2}$ 
being dropped at $10$~kpc. Yet our clouds do not reach terminal velocities, but are successively decelerated
due to continuous mass loss. As demonstrated by models Hc1b13a and Hc1b15b,
the more massive fragments are to be found at lower $z$, 
and -- as discussed above -- the fragments lag behind
the main body of the cloud. This should be observable as a gradient along head-tail clouds or a shift in 
velocity components between warm and cold gas
\citep{2000A&A...357..120B,2001A&A...370L..26B}.

\subsection{Characteristic Disruption Times and Distances}\label{ss:disrupt} 

Our model data allow us to estimate characteristic timescales $\tau_0$ and distances $D_0$ within which
a cloud of a given mass and at a given background halo density will lose its H{\small{I}} content.
This information we then can use to predict whether observed HVCs with distance constraints will
retain any of their H{\small{I}}, or whether they will disintegrate before they reach the disk.
The time series of our models provide us with an opportunity to sample a whole range in cloud 
mass and (for series H) in halo densities. 

\begin{figure} 
  \begin{center} 
  \includegraphics[width=\columnwidth]{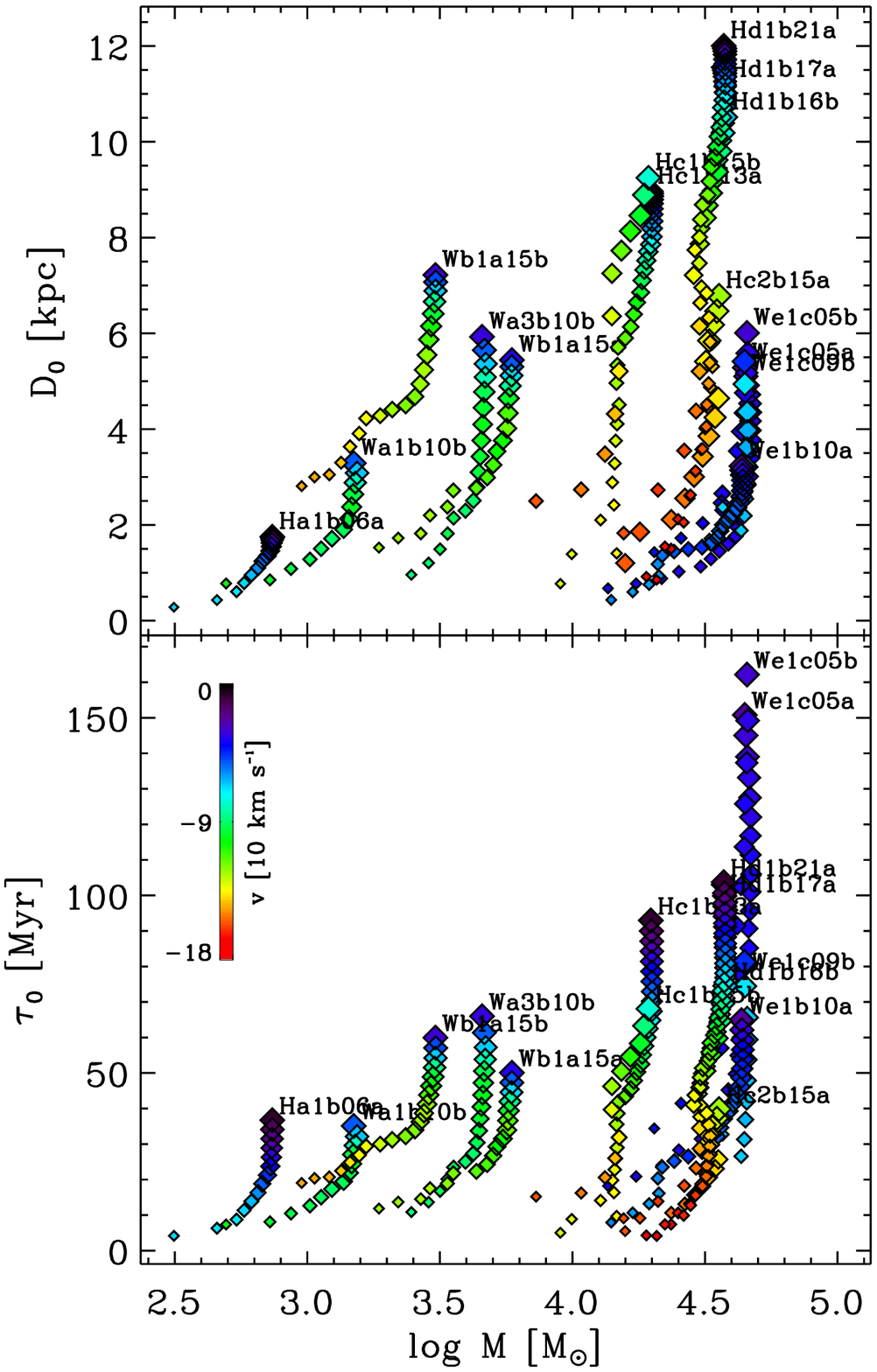} 
  \end{center} 
  \caption{\label{f:charzt}Distance $D_0$ and timescale $\tau_0$ within which a cloud of given mass will  
          lose its total H{\small{I}} content. Each "track" represents the time sequence of one model,
          with each point indicating $D_0$ or $\tau_0$ of the largest fragment
          (see eqs.~[\ref{e:rate_tau},\ref{e:rate_dist}] and text).
          Thus, the disruption rate $dM/dt$ increases over time. Colors denote the 
          cloud velocities, showing that faster clouds are shorter-lived. 
          Symbol sizes indicate model times, with large symbols for early times, and small ones for
          late times. 
          Note that $D_0$ is not a distance above the plane, but a disruption length scale.}
\end{figure}

The disruption time and distance are given by 
\begin{equation}
  \tau_0 \equiv \frac{M}{dM/dt}\label{e:rate_tau}
\end{equation}
and 
\begin{equation}
  D_0 \equiv \frac{M}{dM/dz}\label{e:rate_dist},
\end{equation}
with the cloud mass $M$, the mass loss rate with respect to time $dM/dt$ and distance $dM/dz$. 
The mass range is sampled by taking the differences (here for time)
\begin{equation}
  \frac{dM}{dt}\equiv \frac{M(t)-M_{end}}{t-t_{end}}\label{e:deriv}
\end{equation}
for all model times $t>t_{end}$, where $t_{end}$ is
the smaller of the time at which the cloud has lost $90$\% of its original H{\small{I}} content, 
and the end of the simulation. 
Figure~\ref{f:charzt} summarizes $D_0$ and $\tau_0$ in dependence of the cloud mass, for all models.
Colors indicate the velocity $v(t)$ of the cloud.

Most clouds are disrupted within $10$~kpc and $100$~Myr, with the majority surviving for not longer
than $8$~kpc. Slower clouds survive for longer times and distances, mirroring the lower mass-loss rates,
with clouds at $v_z\gtrsim 100$~km~s$^{-1}$ lasting for at most $8$~kpc. Higher background halo densities
also lead to faster disruption, as expected. 

\begin{figure*}
  \begin{center}
    \includegraphics[width=\textwidth]{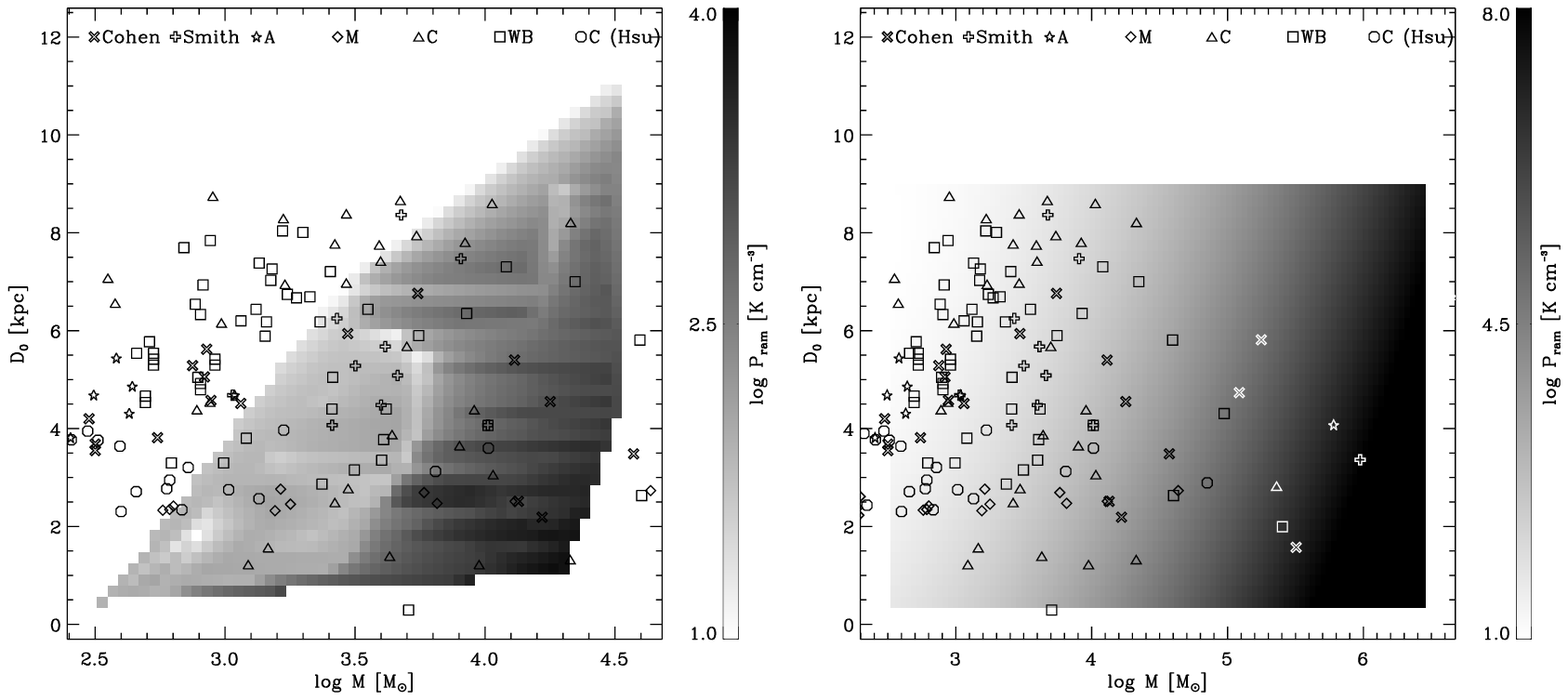}
  \end{center}
  \caption{\label{f:charobs} {\em Left:} Disruption distance $D_0$ derived from models shown
           in Fig.~\ref{f:charzt}, against cloud mass $M$, for a range
           of background halo densities $n_h$ and cloud velocities, combined into a ram pressure 
           $m_Hn_hv^2/k_B$ (see colorbar). A ram pressure of $10^4$~K~cm$^{-3}$ corresponds
           to approximately $100$~km~s$^{-1}$ at a background halo density of $10^{-2}$~cm$^{-3}$.
           Symbols represent the $z$ height above the plane and mass of clouds that are part of HVC 
           complexes with known distance constraints (see Tab.~\ref{t:hvcdist}).
           {\em Right:} Similar to left panel, but using a two-dimensional quadratic fit to the ram
           pressure surface, and extrapolating
           to higher masses to estimate the disruption distances for more massive clouds.
           Except for the main body of Complex C (at $\approx 3\times 10^6$~M$_\odot$), all clouds of the observational
           comparison sample are shown.} 
\end{figure*}

Figure~\ref{f:charobs} can be used to predict the disruption distances $D_0$ for observed HVCs, given 
their mass, the halo background density $n_h$ and the cloud velocity $v$. The latter two have been combined
into the ram pressure (in K~cm$^{-3}$). The left panel of the Figure
shows the disruption distance $D_0$ at a given mass and (in grayscale) ram pressure of all models, derived
from Figure~\ref{f:charzt}. Overplotted are the masses and distances above the plane $z$ for a sample
of HVCs taken from the extended catalogue by \citet{1991A&A...250..509W} with $\sim30$\arcmin~resolution, 
and from a catalog of the clouds at the tail of Complex C as observed with Arecibo by 
Hsu et al. (2009; 3.5\arcmin~resolution). 
We selected only clouds of HVC complexes with known distance constraints (see Tab~\ref{t:hvcdist}).
Masses have been calculated from the distance constraints and the total flux per cloud as
$M_{HI} [\mbox{M}_\odot] = 2.36\times 10^{-1} D_{kpc}^2 F$ [Jy~km~s$^{-1}$].

\begin{deluxetable}{lccc}
\tablewidth{0pt}
\tablecaption{Distance Constraints\label{t:hvcdist}}
\tablehead{\colhead{Complex}&
           \colhead{$D$ [kpc]}&
           \colhead{$z$ [kpc]}&
           \colhead{reference}}
\startdata
ACHV/Cohen Strm & $5.0$-$11.7$       & $0.9$-$9.4$  & (1)\\
GCP/Smith Cld  & $9.8$-$15.1$       & $2.6$-$10.1$ & (1)\\
A              & $4$-$10$           & $2.2$-$7.7$  & (2)\\
M                & $1.5$-$4.4$        & $1.1$-$4.0$  & (3)\\
C                & $10.0\pm 2.5$      & $0.9$-$11.0$ & (4,5)\\
WB                & $8.8^{+2.3}_{-1.3}$& $1.7$-$10.2$ & (6) 
\enddata
\tablecomments{1st column: complex name,
2nd column: heliocentric distance,
3rd column: distance towards Galactic plane,
4th column: reference: (1) \citet{2008ApJ...672..298W}, (2) \citet{1999Natur.400..138V},
(3) \citet{1993ApJ...416L..29D}, (4) \citet{2008ApJ...684..364T}, 
(5) \citet{2007ApJ...670L.113W}, (6) \citet{2006ApJ...638L..97T}.
The ranges in $z$ have been estimated using fiducial distances $D$ of $8.4$, $12.5$, $7.0$, and
$3.0$~kpc for complexes with $D$-ranges given, and then accounting for the full extent in Galactic latitude.
}
\end{deluxetable}

Clearly, model clouds with higher mass tend to travel farther distances until their H{\small{I}} content
is completely gone -- indicated by the upper envelope of the grayscale map. This effect is offset by
an increased ram pressure -- an increase by a factor of $10$ reduces the disruption length scale $D_0$ by
approximately the same factor. 

Figure~\ref{f:charobs} shows that many of the observed HVCs will not make it to the disk in cold form.
The left panel is restricted to the mass range sampled by our model HVCs.
Clouds located above the gray shaded region will clearly disrupt before making it to the disk, while clouds
within the Gray shaded region may make it depending on the exact halo density they are moving through
and their velocity. 
Yet since more than $90$\% of the mass in the HVC complexes with distance constraints  
resides in clouds of large masses (as catalogued by \citealp{1991A&A...250..509W}), the right panel of Figure~\ref{f:charobs} shows
an extrapolation of the ram pressure map to larger masses. The extrapolation has been calculated
by a two-dimensional quadratic fit to the ram pressure surface shown in the left panel. Although
this approach is rather crude, it demonstrates the salient correlation between $D_0$, $M$ and  $P_{ram}$.
While the smaller HVCs will most likely be disrupted on their way to the disk, the
situation is less clear for the main body of the larger complexes at these distances.  Yet it should 
be noted that when higher resolution catalogs are used, these massive clouds are often broken up into
smaller clouds (\citealp{2002AJ....123..873P}; Fig.~1 of \citealp{2004ASSL..312..145S}). In other words, the fact that
the substructure in HVC complexes tends to be underestimated renders the predicted disruption distances
of Figure~\ref{f:charobs} as upper limits.

\begin{figure}
    \includegraphics[width=\columnwidth]{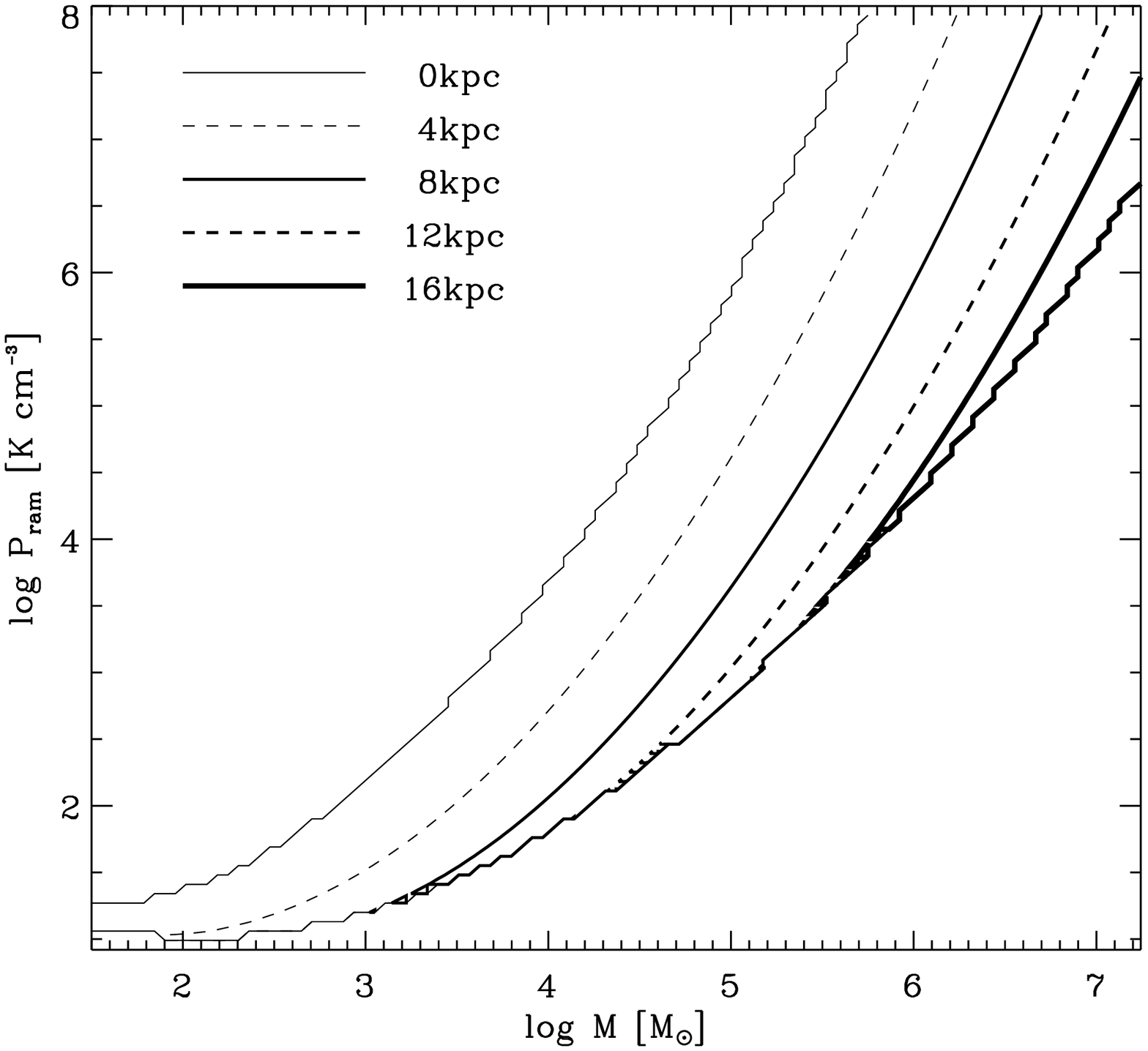}
  \caption{\label{f:easyd}Contours of the disruption length scale $D_0$ in kpc, as a 
           function of the cloud mass $M$ and the ram pressure $P$. A ram pressure of $10^4$~K~cm$^{-3}$ corresponds
           to approximately $100$~km~s$^{-1}$ at a background halo density of $10^{-2}$~cm$^{-3}$.
           The range of applicability of equation~(\ref{e:d0form}) is indicated by the $0$-line towards higher
           $P$ and lower $M$, and the thick line towards lower $P$ and higher $M$ (i.e. $D_0 \lesssim 16$~kpc).}
\end{figure}

Figure~\ref{f:easyd} offers a straight-forward way\footnote{We thank the referee for the suggestion.} 
to read off the disruption length scale $D_0$ as a function
of $\log M$ and the ram pressure $\log P$. This has been derived by inverting the two-dimensional quadratic fit,
yielding 
\begin{equation}
\begin{split}
D_0 =   3.481\log M - \Bigl(&-165.2+77.16\log M\\ 
                            &-18.54(\log M)^2+86.89\log P\Bigr)^{1/2}\\ 
                            &\phantom{-}\mbox{ [kpc]},\\   
\end{split}
\label{e:d0form}
\end{equation}
where $M$ is given in $M_\odot$, and $P$ in K~cm$^{-3}$.
Since the quadratic form represents the data reasonably well only in a limited range of parameters, 
equation~(\ref{e:d0form}) is limited to the range in $M$ and $P$ indicated by the outer contours in Figure~\ref{f:easyd}. 
Within that range, equation~(\ref{e:d0form})
is accurate to approximately $10$\%.
Below the thick line towards high masses and low pressures (for $D_0$ exceeding $16$~kpc), 
equation~(\ref{e:d0form}) breaks down.
Still, it allows us to read off the approximate disruption length scale $D_0$ for physically reasonable
values of $M$ and $P$.

\subsection{Resolution issues}\label{ss:resoissue}
Numerical resolution is a main concern when modeling the disruption of clouds by dynamical
instabilities. In their two-dimensional study of cloud disruption by passing shocks, 
\citet{1994ApJ...420..213K} quote a resolution of approximately $100$ cells per cloud radius (or $R100$)
to fully resolve the flow dynamics around the cloud.
Yet \citet{2000Icar..146..387K}  noted that since the dynamical instabilities lead to non-linearities
and turbulence, detailed convergence might be unfeasible (see also \citealp{1994ApJ...434L..33M}). 
They argue for statistical
convergence (or convergence of integrated quantities) at $\approx R100$.
In a similar vein, \citet{1997Icar..126..138R}
argued in the context of the disruption of comets in planetary atmospheres that $R25$ is 
sufficient to converge on the penetration depth within a scale height. 

Checking the lower panels of Figure~\ref{f:timesum}, it is obvious that our models 
will have a hard time meeting the resolution criterion of $R100$ for all of the cloud fragments, especially at later times
when the main body of the cloud will develop a strong tail, reducing its cross section
and thus possibly suppressing Rayleigh-Taylor instabilities \citep{1997Icar..126..138R,2000Icar..146..387K}.
Clearly, the cloud fragments will be dynamically underresolved, an effect which is
exacerbated in the models of the H-series, due to the increasing external pressure.
The smallest cloudlets might be dominated by evaporation (e.g. \citealp{2008ApJ...680..276S}
in the context of the Magellanic Stream). 

\begin{figure} 
  \begin{center} 
  \includegraphics[width=\columnwidth]{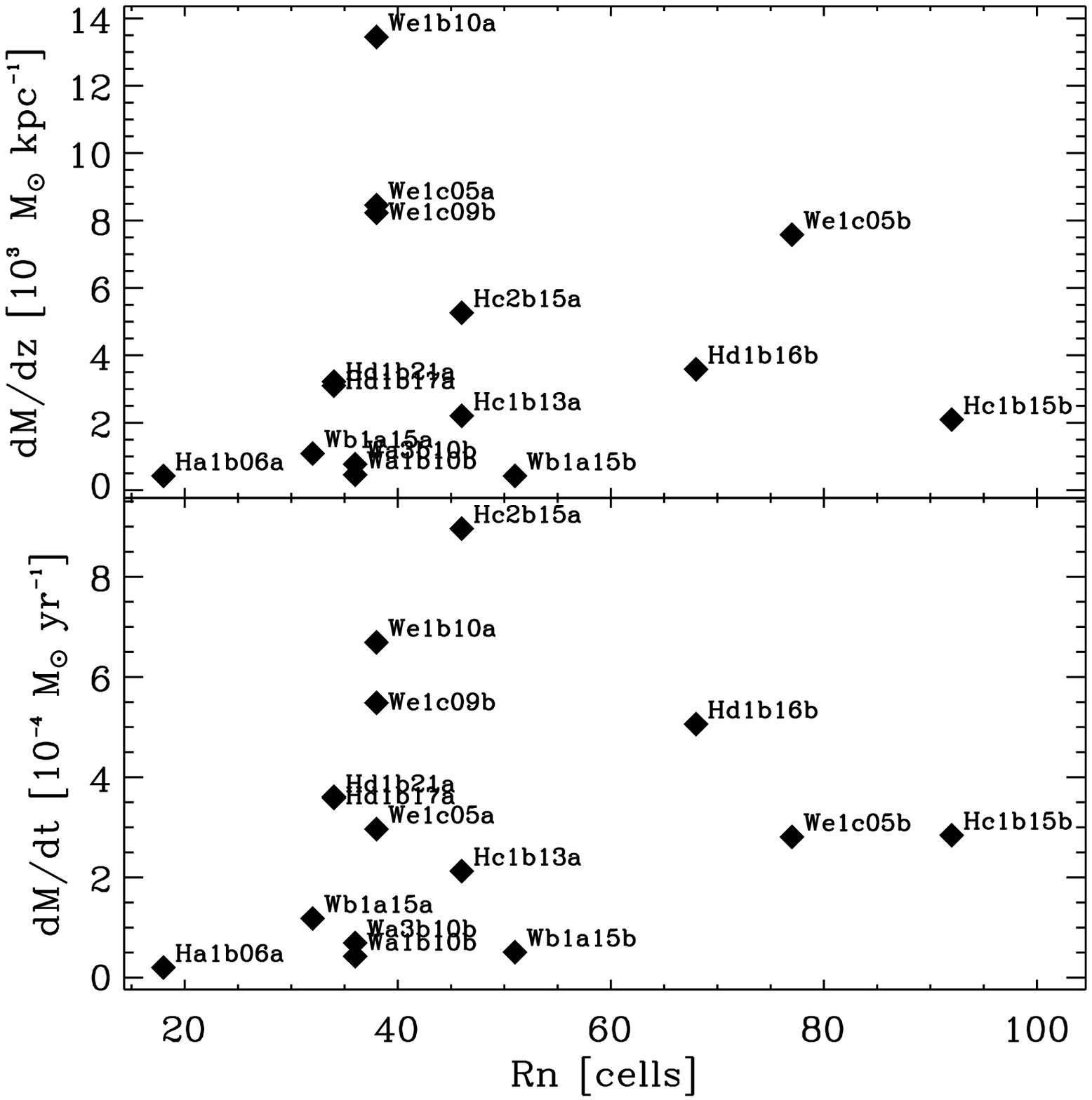} 
  \end{center} 
  \caption{\label{f:reso}Mass loss rate in terms of time (bottom) and distance traveled (top),
           against the number of resolution elements per radius $Rn$. 
           Ignoring the clearly
           underresolved model Ha1b06a, there is no trend of mass loss rates with resolution.}
\end{figure}

Figure~\ref{f:reso} helps us to assess the effect of $Rn$ on the mass loss rate in terms
of time and distance traveled (eqs.~[\ref{e:rate_tau},\ref{e:rate_dist}]).
The mass loss rates will be correlated with the strength of the dynamical instabilities
disrupting the cloud (e.g. \citealp{1982MNRAS.198.1007N}; \citealp{2000Icar..146..387K}), 
which in turn will be suppressed in underresolved models.
No correlation between mass loss rates and resolution is discernible, if 
we ignore the clearly underresolved model Ha1b06a. Although this clearly does not mean
that the models are fully resolved, we are confident that they are sufficiently resolved
for our purposes.

\subsection{Differences to Previous Numerical Work}\label{ss:previous}
The wind-tunnel experiments (series W) are conceptually similar to the models discussed
by \citet{2001ApJ...555L..95Q}. Consistent with their claim,
we do not notice a substantial difference in disruption timescales 
between models with and those without cooling\footnote{Although we ran a few comparison
models without cooling, we decided not to discuss them in more detail, since their restricted physics
render them irrelevant for our purposes.}.
However, for series H,  the comparison models
without cooling showed substantially shorter disruption times and distances, since the 
cloud suffers adiabatic compression and heating due to the increasing background pressure.
Thus, the initially cold material is rapidly heated, and thus "lost" (i.e. not identified
as H{\small{I}} any more). In that sense, cooling
is relevant for the longer survival of the clouds of series H, and it is obviously indispensable
for the recooling observed in e.g. model Hc1b13a.

\citet{2004JKAS...37..233S} studied the impact of HVCs on the Galactic disk using three-dimensional isothermal
magneto-hydrodynamical (MHD) models. Focusing on the physics of the impact, they had the HVC start at $z=2.5$~kpc above
the plane. Thus, conclusions on the possibly stabilizing effects of magnetic fields on HVCs over long distances
could not be addressed. 

A related problem albeit in a different physical regime was considered by \citet{2007MNRAS.378..662R} who studied 
the disruption
of buoyant bubbles driven by AGN feedback in galaxy clusters. They found that
tangled magnetic fields can prevent the disruption of the bubbles if the coherence length of the field is larger than
the bubble size, i.e. if the field is only weakly tangled with respect to the bubble scale. 
Otherwise the bubbles are quickly shredded.   

\citet{2007ApJ...670L.109B} focus on the origin of the H$\alpha$ emission along the Magellanic Stream
and explain it by a model of successively shocked material as it is ablated from clouds in the Stream
predominantly by Kelvin-Helmholtz instabilities
(see however \citealp{2000ApJ...529L..81M}).
Since they are interested in the dynamics of the Stream's gas, they start with a two-phase medium, with
the cold phase having a fractal distribution. Their numerical method includes non-equilibrium cooling.
They speculate that much of the gas accretion necessary to fuel Galactic star formation happens in the 
warm ionized phase, largely of material shorn off the Magellanic Stream (see also \citealp{2008arXiv0811.2467B}).
Clearly, as shown above, H{\small{I}} clouds originating at the distance of the Magellanic Stream will not survive
the trip to the disk.

%
%
\section{Discussion} 
 
\subsection{Interpretation of Morphologies}\label{ss:intermorph}
The model clouds usually develop a pronounced tail, visible not only in the maps (Fig.~\ref{f:acviews}),
but also in the velocity lags of the smaller H{\small{I}} fragments (Figs.~\ref{f:timesum},\ref{f:distsum}).
The lag is consistent with the velocity gradients and/or shifts between warm and cold gas components in
observed HVCs (\citealp{2000A&A...357..120B,2001A&A...370L..26B}; see also \citealp{2005A&A...432..937W}) and should
be examined further in higher resolution surveys with Parkes and Arecibo.
The head-tail structure will be most pronounced when the trajectory of the cloud is perpendicular
to the line-of-sight.  HVCs that appear round may show a kinematic offset in their cold and warm 
components providing additional evidence of clouds being destroyed by the surrounding halo medium.   
In the future it should be possible to create a grid of the typical length/velocity offset expected 
for clouds at different stages of destruction and viewing angles.

\subsection{Interpretation of Timescales and Distances}\label{ss:interdist}
From Figure~\ref{f:charzt} we see that the majority of our 
model clouds will be disrupted within $\lesssim 8$-$10$~kpc.
However, various physical and numerical issues will affect our estimates.

\subsubsection{Effects Reducing $\tau_0$ and $D_0$}\label{sss:reduce}
The following effects will render our estimates as upper limits.
\paragraph{(1)} The actual halo density may be larger than our maximum of $n_h = 3\times 10^{-4}$~cm$^{-3}$. 
\citet{2008PASA...25..184G} set limits on the halo  
density at $z>5$~kpc using the $6$ measurements they have that fit that  
criterion.  They find a $3\sigma$ upper limit of $n_h < 7.6\times 10^{-4}$~cm$^{-3}$.
This type of limit is consistent with other work based on pulsar dispersion measures
(e.g. \citealp{2006ApJ...637..333H}). 
The majority of the constraints on the halo density are  
integrated with assumptions on the temperature and structure of the  
halo.  To explain observations of O{\small{VI}} absorption, \citet{2003ApJS..146..165S} propose 
densities $\lesssim 10^{-4}$ down to $10^{-5}$~cm$^{-3}$.  
\citet{2007ApJ...656..907P} find consistency with a  
density of a few $10^{-4}$~cm$^{-3}$ at about $10$~kpc, and \citet{2009arXiv0901.4975G}  
estimate that a halo density $>10^{-4}$~cm$^{-3}$ is likely at these types  
of distances to explain dwarf spheroidal stripping.
\citet{2007ApJ...669..990B} and \citet{2002AAS...201.4705S}
combine X-ray absorption and emission  
measures and find a density of  $9\times 10^{-4}$~cm$^{-3}$ for a lengthscale  
of $20$~kpc. See also \citet{2007ARA&A..45..221B} for a discussion.

\paragraph{(2)} The inflow gas density in the W-series is constant, i.e. we do not take into account the 
increase of the gas density along the trajectory of the cloud towards smaller $z$ for those models. 

\paragraph{(3)} The velocity of our clouds gradually increases to the velocity listed in Table~\ref{t:param}, 
so any initial velocity from the origin of the cloud -- e.g. the motion of the Milky Way satellite -- 
is not included.

\paragraph{(4)} Resolution effects -- in the H-series exacerbated by the increased cooling due to the
rising external pressure -- tend to work in favor of faster disruption: 
under-resolved clouds tend to take longer to disrupt
(\citealp{1997Icar..126..138R}; \citealp{2000Icar..146..387K}; see also Fig.~\ref{f:reso}).

\paragraph{(5)} Assuming that the cooling timescale does not substantially decrease during
the lifetime of the cloud, breaking the cloud into successively smaller fragments should render
$\tau_c/\tau_d>1$ eventually, i.e. the fragments should be more likely to disrupt due to 
dynamical instabilities, and eventually through evaporation \citep{1977ApJ...211..135C}.

\paragraph{(6)} The substructure in HVCs and complexes tends to be underestimated
(see \citealp{2002AJ....123..873P}, and Fig.~1 of \citealp{2004ASSL..312..145S}). 
Existing substructure in HVCs (whether
imprinted during their formation or already due to the interaction with the background
medium) will accelerate the disruption. The presence of substructure is equivalent to
a gas volume filling factor $f_V<1$, i.e. some part of the 
cloud volume is at densities
lower than the mean density (and if $f_V\ll 1$, a large part). Since the dynamical instabilities responsible for
fragmentation have shorter growth timescales at lower density contrasts, a small volume
filling factor promotes cloud disruption. Moreover, the sound speed will be larger 
and the cooling time will be longer in a large part of the cloud volume, resulting in
a dominance of dynamical over thermal effects. 
 
\subsubsection{Effects Increasing $\tau_0$ and $D_0$}\label{sss:increase}
There are also mechanisms that will potentially increase the lifetimes of the halo clouds.
Magnetic fields can suppress the dynamical instabilities 
\citep{1961hhs..book.....C}, thus possibly leading to longer cloud lifetimes. Two-dimensional  
numerical models seem to support this notion (\citealp{2002A&A...391..713K}; see also \citealp{2008ApJ...678..234P}), 
however, the dominant  
instabilities allow interchange modes in three dimensions, possibly with growthrates above the 
hydrodynamical ones (e.g. \citealp{2007ApJ...671.1726S} for the  
Rayleigh-Taylor instability). Currently available three-dimensional models of HVCs including 
magnetic fields \citep{2004JKAS...37..233S} are limited to clouds starting at $z=2.5$~kpc above the plane 
and thus cannot address the question of cloud disruption over longer distances.
Magnetic field estimates in HVCs are scarce -- an estimate of $11.4\pm 2.4\mu$G in a single HVC
\citep{1991A&A...245L..17K} could not be confirmed (\citealp{1995ApJ...451..645V}: $-0.1\pm 1.8\mu$G for the same
object). Indirect estimates based on cosmic ray confinement quote a few $\mu$G 
(\citealp{1990A&A...239...57V} for the Galaxy, and \citealp{1994A&A...292..409B} for the starburst 
galaxy NGC 253; see also \citealp{2008PASA...25..184G}). 
With typical halo gas parameters, this would result in a plasma $\beta$ of order unity, i.e. the fields 
might not be dominant. Clearly, the effects of magnetic fields need to be explored further.
 
\citet{2007A&A...472..141V,2007A&A...475..251V} argue on the basis of high-resolution two-dimensional 
models of HVCs that heat conduction should play a substantial role in stabilizing the  
clouds against disruption by introducing a smooth transition layer between the hot and cold gas. 
Yet the effect seems to extend only the lifetime of a uniform-density cloud by a factor of two, 
while centrally condensed clouds (see Figs. 4a, 6a and 9a of \citealp{2007A&A...472..141V}) are not
strongly affected, presumably because they already possess a smooth transition layer, intrinsically
weakening the dynamical instabilities. 
 
Viscosity could also play a role in stabilizing the cloud against dynamical instabilities
(e.g. \citealp{1961hhs..book.....C}) by introducing a finite shear layer and thus reducing
the relative velocities essential for a rapid (and unconditional) growth of the Kelvin-Helmholtz 
instability. Since viscosity is included in our numerical scheme at (resolution-limited) Reynolds
numbers smaller than those expected for the Galactic halo, its stabilizing effect is actually 
over-estimated in our models.

Given the distance constraints on the HVCs discussed here, our model HVCs are not massive enough
to be self-gravitating, nor are they expected to be dominated by dark matter 
(e.g. \citealp{2004ASSL..312...25W}). Thus, we are neglecting the otherwise stabilizing effect
of self-gravity (e.g. \citealp{1993ApJ...407..588M}) and of external potentials (e.g. \citealp{2000ApJ...538..559M}).

\subsection{Consequences for Feeding the Galactic Disc}\label{ss:feeding}
Taking our estimated disruption timescales and distances, we find that clouds at $10$~kpc or less above the disk
with HI masses $< 10^{4.5}$~M$_\odot$ are highly unlikely to make it to the disk in the form of neutral hydrogen clouds.  
Because of the limited mass range sampled ($3\times 10^2<M_{HI}<4.4\times 10^{4}~$M$_\odot$),
we can only extrapolate our model data to predict the evolution of larger HVCs such as Complex~C
(at $\approx 10^6$~M$_\odot$; see Figure 6).  These large complexes are often seen to consist of numerous
smaller clouds when observed at higher resolution (\citealp{2004ASSL..312..145S}; \citealp{2002AJ....123..873P}; Hsu et al. 2009), so
much of their mass may also have difficulty making it to the disk in cold form.   
 
The lack of surviving clouds has implications for the accretion of gas from satellites.
The Magellanic Stream is 
an example of a larger complex breaking up into smaller clouds at the tip 
\citep{2008ApJ...680..276S} and via the numerous small clouds found along its length 
\citep{2003ApJ...586..170P}. Given the distance of the Magellanic
Clouds ($\sim 60$ kpc) the majority of the Stream is unlikely to make it to the disk in the form 
of cold HI clouds (see also \citealp{2007ApJ...670L.109B}, \citealp{2008arXiv0811.2467B}). 
Excluding the Magellanic System, \citet{2009arXiv0901.4975G} find that all satellites within 270 kpc of the 
Milky Way or Andromeda are devoid of HI gas. This gas was most likely primarily stripped
at the satellite galaxy's perigalacticon, but this still means the gas was stripped from satellites at
typical distances of $20$-$120$ kpc.   Given these distances the majority of the gas stripped from satellites
is unlikely to make it to the galaxy in cold form.  If the fate of satellite galaxy gas is to ultimately 
provide the Galaxy with star formation
fuel, it must first be integrated into the previously existing
warm/hot halo medium.

The destruction of cold clouds in the halo results in a complex 
multi-phase halo of warm gas intermingled with the hot and cold gas, as also evident through 
the absorption line observations \citep{2003ApJS..146..165S,2008A&A...487..583B} and
pulsar dispersion measurements (\citealp{2008PASA...25..184G}, although given the pulsar 
distances, most of the material they see should be part of a thick disk component. 
See also \S\ref{sss:reduce}).
Even if a large fraction of the clouds do not reach the 
disk in the form of neutral hydrogen, they still are over-dense with respect to the 
background medium, and thus sink towards the plane (see also \citealp{2007ApJ...670L.109B}). 
One might speculate whether such remnants contribute to the population of warm ionized medium 
(WIM) clouds in the hot coronal gas (\citealp{1993AIPC..278..156R}; \citealp{2008PASA...25..184G}). 
Once they are within $1$--$2$~kpc of the Galactic disk, the gas flows driven by feedback
from massive star forming regions (Galactic fountain, see e.g. \citealp{2004ASSL..312..341B} 
for a discussion)
will most likely lead to compressions and recooling, eventually integrating
the former HVC material into the disk as new star formation fuel. 
This could be easily tested by placing tracer particles in  models of 
stratified Galactic disks including
supernova feedback, such as by \citet{2006ApJ...653.1266J}.

The WIM phase is difficult to trace in our simulations, as it moves at 
slower velocities than the H{\small{I}} clouds due to 
being more diffuse, and thus leaves the simulation domain rather quickly (within $39$~Myr at
$50$~km~s$^{-1}$). Yet in models Hc1b13a and Hc1b15b (Fig.~\ref{f:distsum}) we were able to follow
the cloud evolution long enough to observe its nearly complete deceleration close to the plane, and
the subsequent recooling of stripped material. These models suggest that warm ionized gas stripped
from a HVC can recool at low $z$ (even without gas dynamics within the disk, as mentioned above), and
thus re-form cold H{\small{I}} clouds. Because of their velocities, they would be classified as 
intermediate or low-velocity clouds (IVCs, LVCs).  These LVCs may be similar in nature to the discrete 
clouds recently detected in multiple directions by several HI surveys 
(\citealp{2002ApJ...580L..47L}; \citealp{2006ApJ...653.1210S}; \citealp{2008ApJ...688..290F}). 
These observed clouds are small and cold and are most likely embedded in the WIM extending
$\sim3$ kpc above the plane.   If they are a mixture of accreting gas from the halo and fountain material, 
the LVCs should have sub-solar metallicities and may themselves show head-tail cloud structures as they sink 
into the Galactic disk.  

%
%
\section{Summary} 
 
Motivated by the observed head-tail structure of HVCs  
(e.g. \citealp{2000A&A...357..120B}; Putman et al., in prep.; Fig.~\ref{f:acviews})
and their potential to constrain the properties of the diffuse Galactic halo,
we study numerically the disruption of  HVCs during their passage through the Galactic halo.
Our experiments address the two possible extremes of HVC trajectories, namely (a) assuming a 
constant background halo density and pressure, and (b) assuming an isothermal, hydrostatic halo
with the cloud falling straight towards the disk. These are our findings:

\paragraph{Cloud Morphology and Evolution}
In agreement with previous numerical studies, the clouds develop a pronounced head-tail structure
due to the interaction with the background halo medium (Fig.~\ref{f:acviews}). 
Cooling does not necessarily stabilize the cloud, especially when the dynamical timescales are shorter than
the cooling timescale (Fig.~\ref{f:timesum}). The velocity difference between head and tail is 
consistent with observed lags (Fig.~\ref{f:distsum}), and we suggest that observed radial cloud velocities
should be correlated with the shape of the cloud and the head-tail lag.

\paragraph{Cloud Disruption and Characteristic Distances}
The limited numerical resolution and our conservative parameter choice (\S\ref{sss:reduce}) renders
the estimated cloud disruption timescales and distances as upper limits.  Therefore, HVCs with
H{\small{I}} masses $< 10^{4.5}$~M$_\odot$ will lose their H{\small{I}} content within $10$~kpc or less
(Fig.~\ref{f:charobs}). Extrapolating to more massive complexes (Fig.~\ref{f:easyd}), extreme ram pressures 
(e.g. $n_h>10^{-2}$~cm$^{-3}$ and $v>100$~km~s$^{-1}$) or the fact that these complexes have
abundant substructure (see \S\ref{sss:reduce}) could lead to much of their H{\small{I}} mass being lost before
reaching the disk.
Material from the Magellanic Stream and other dwarf galaxies is not expected to survive the trip to the disk in the form of
H{\small{I}} clouds.

\paragraph{The Lower Halo and Upper Disk: Fueling Star Formation}
Warm ionized material stripped from the H{\small{I}} clouds is still falling towards the disk (albeit
at lower velocities). This material might constitute some of the extended layer of warm ionized gas
(\citealp{1993AIPC..278..156R}; \citealp{2008PASA...25..184G}). 
Some of this material can actually recool and thus re-form H{\small{I}} cloudlets
close to the disk. Because of their low velocities, such clouds would be identified as IVCs or LVCs.
We argue that the HVC remnants still could merge with the Galactic disk as WIM clouds or via recooling close to the disk,
and eventually fuel the star formation in our Galaxy via a warm cosmic rain.

\acknowledgements  
We are grateful to the referee, R.~Benjamin, for a very detailed,
critical and thorough report, and for all the prying questions by which
he made us expand and strengthen our presentation considerably. 
Comments by J.~Bullock, J.~Bland-Hawthorn, T.~Kaufmann and A.~Maller on an earlier version of
the manuscript are highly appreciated. Computations were
performed at the National Center for Supercomputing Applications 
(AST 060034) and on the local PC cluster Star, perfectly maintained and 
administered by J.~Hallum \& J.~Billings. We acknowledge support 
from the Research Corporation and from NSF grants 
AST~0707597, AST~0748334 and AST~0807305. The manuscript was revised substantially
at the Kavli Institute for Theoretical Physics (KITP) during the last week 
of the "Building the Galaxy"-Program,
supported by NSF PHY05-51164.
Thanks go to J.~Kollmeier and the KITP for the invitation, and to 
J.~Peek and J.~Bland-Hawthorn for discussions. 
This work has made use of NASA's Astrophysics 
Data System. Last but not least, F.H. thanks {\em S}.~Putman for
many inspirational runs. 
 
%
%
 
\bibliographystyle{apj} 
\bibliography{./references}

\end{document}